\documentclass[smallextended,final]{svjour3}
\usepackage[utf8]{inputenc}
\usepackage[greek,english]{babel}
\usepackage{graphicx}
\usepackage{textcomp}
\usepackage{color}
\usepackage{amssymb,amsmath}
\usepackage{hyperref}
\usepackage{times}
\usepackage[authoryear,round]{natbib}
\usepackage{multirow}
\usepackage{arydshln}
\usepackage{enumitem}

\title{Science and illusions}
\author{Luigi Scorzato}
\institute{Present address: Accenture AG, Digital, Big Data Analytics. 20 Rue de Pr{\'e}-Bois CH-1215 Gen{\`e}ve. 
\email{luigi@scorzato.it}}
\date{}

\begin{document}

\newtheorem{defn}{Definition}
\newtheorem{post}{Postulate}
\newtheorem{rmrk}{Remark}

\maketitle

\begin{abstract}
It is mostly agreed that Popper's criterion of falsifiability fails to provide a useful demarcation between science
and pseudo-science, because ad-hoc assumptions are always able to save any theory that conflicts with the empirical
data (a.k.a. the Duhem-Quine problem), and a characterization of ad-hoc assumptions is lacking.  Moreover, adding
some testable predictions is not very difficult.  It should be emphasized that these problems do not simply make
Popper's demarcation {\em approximate} (if it were so, all our problems would be solved!), they make it {\em
  totally useless}.  More in general, no philosophical criterion of demarcation is presently able to rule out even
some of the most blatant cases of pseudo-science, not even approximatively (in any well defined sense of
approximation).  This is in sharp contrast with our firm belief that some theories are clearly not scientific.
Where does this belief come from?  In this paper I argue that it is necessary and possible to recognize the notion
of syntactic simplicity that is able to tell the difference between empirically equivalent scientific and
non-scientific theories, with a confidence that is adequate to many important practical purposes, and it fully
agrees with the judgments generally held in the scientific community.
\end{abstract}

\section{Introduction --- sophisticated pseudo-science}


This paper proposes an approximate solution to the classical demarcation problem, which is satisfactory to most
practical purposes\footnote{In \citet{Scorzato-paper2, Scorzato-silfs14} the same framework is used to define the
  concept of scientific progress.  To keep this paper self-contained, all the relevant definitions are repeated.}.
To explain why no present solution is satisfactory --- not even approximatively --- I start from an example, which
bears the essential features of all today most threatening cases of pseudo-science.  Imagine the following story.
In June 2015 Prof.~Lucky Furbetto proposed a new theory of gravity, that is identical to Einstein's theory, except
that Furbetto assumed that one of the constants of the theory (say $G$) is actually not constant, but it depends on
time (in the Earth's frame). In other words, he replaced Einstein's constant $G$ with some function $G(t)$.  The
function $G(t)$ is such that no difference with Einstein's original theory could be detected in practice up to now,
but it will have dramatic effects starting from Decembre 31st 2017.  To support his theory, Furbetto also offered a
prediction: by adding more assumptions that, he claims, are somehow justified by his $G(t)$, he was able to predict
three numbers that were indeed extracted at the lottery of July 18th 2015\footnote{Other people had made similar
  attempts before, but Furbetto claims that they are totally unrelated to him, and indeed their theories are
  completely different.  Nobody ever blamed Einstein for the wrong predictions made by other people...}.  A second
lottery prediction is foreseen for the summer 2025, but now Furbetto claims that we should all get ready to face
the terrible disasters of 31 December 2017.

Intuitively, we know that Furbetto's theory deserves no serious consideration.  But why, {\em exactly}?  If we
can't answer this question, how can we hope to convincingly rebutting more seducing cases of pseudo-science?


Unfortunately, this question is subtle because Furbetto's theory can't be criticized on purely empirical grounds:
it agrees with all known experiments as much as our best theories.  It even got a very non-trivial and impressive
prediction right, that no other theory could match.  What could we wish more from the empirical point of view?  

If we observe that the right lottery numbers may be derived from a modification of his ad-hoc assumptions without
using G(t) at all, Furbetto may rightly argue that even the bending of light can be derived by some other ad-hoc
assumptions without using Einstein's theory at all. But in the way Furbetto derived the lottery numbers, G(t) was
actually essential \footnote{This can be achieved, e.g. by assuming: {\tt (lottery-outcome = $(7, 13, 42)$ AND
    $G(t) = 1+exp[t/{\sigma}]$) OR (lottery-outcome = $(3, 18, 51)$ AND $G(t) =$ const)}.}.  Actually, Furbetto is
right to claim that he never chenged his assumptions, as opposed to Einstein, who first took away and then added
again a cosmological constant to match the empirical evidence...

If we argue that there is no {\em reason} to postulate a time dependence on $G$, nor to postulate a sudden change
on December 31st 2017, Furbetto can rightly request us to specify which {\em kind} of reason do we want: there is
no {\em compelling} reason to assume it constant either. On the other hand, he can provide many non-compelling but
very inspiring narratives for his choice.  What do we want exactly?  In conclusion, both from the empirical and the
logical points of view there is very little we can do to reveal any objective weakness of Furbetto's theory.

Popper's criterion offers no help either\footnote{For the same reasons, also Testability \citep{SoberTest} can't
  tell what's wrong with Furbetto's theory.  Indeed, it is not difficult, in general, to elude both falsifiability
  and testability by means of auxiliary assumptions and small, cunning modifications of a theory.}: Furbetto's
theory is as falsifiable as our best theories.  The first prediction was amazingly correct, all other past events
are described correctly, and there are more predictions expected for the end of 2017 and summer 2025.  What else do
we need before taking it seriously?

Even less convincing are those demarcation criteria that actually rely --- more or less explicitly --- on the
opinion of the majority of some scientific community, or on some weird behavior of the proponents of a theory.  To
be clear, Furbetto is an exquisite person, with a vast culture, and he has a lot of friends who have PhDs and think
he is right.  How should that matter?

An interesting view is the proposal to treat science as a family resemblance \citep{DupreDisorder}.  But, which
traits are {\em relevant} to judge such resemblance \citep{PigliucciDem}?  Indeed, in many respects, Furbetto's
theory could be seen as a twin brother of Einstein's theory of gravity.


The reason why no serious scientist needs to wait for December 31st 2017 to declare Furbetto's theory a hoax is not
based on logical and empirical considerations alone, but it is not too difficult to express either: the reason is
that there are {\em simpler} explanations for all the phenomena described by Furbetto's theory.  In fact, assuming
a constant $G$ is simpler than any time dependent $G(t)$, and assuming directly the three lottery numbers, without
much ado, is certainly simpler than Furbetto's attempt to relate them to his $G(t)$, since, anyway, he is not able
to deduce them from his theory of gravity {\em without further assumptions}.  This argument to dismiss Furbetto's
theory is actually fully consistent with a famous definition of science given by Einstein: {\em The grand aim of
  all science is to cover the greatest number of empirical facts by logical deduction from the smallest number of
  hypotheses or axioms\footnote{Quoted in Life Magazine, January 9,1950.}}.


However, without a rule to count the {\em number of hypotheses or axioms}, the previous elegant sentence is
meaningless.  In fact, Furbetto can claim that he can express his theory with just one hypothesis: $\Xi=0$, where
the symbol $\Xi$ summarizes all the equations involved\footnote{Alternatively, $\Xi$ may represent a G{\"o}del
  number that describes the whole system.  I thank Elliott Sober for suggesting this latter representation.}.  On
the basis of this, Furbetto can easily claim that also the simplicity of his theory can't be beaten, since it uses
just one symbol $\Xi$.

There is no chance to prove Furbetto wrong unless we understand --- first of all --- what's wrong with the $\Xi$
formulation.  In fact, if Furbetto claims to be able to interpret $\Xi$ --- and we have no clear argument to refute
that claim --- and if the theory that states just $\Xi=0$ is empirically accurate --- as it certainly is, by
construction --- how could we regard that theory as anything less than optimal by {\em any} respect?


So, can Furbetto really {\em interpret} $\Xi$?  Assessing this is a delicate issue, which is neither trivial nor
hopeless.  But, sadly, in recent decades, both philosophers and scientists have failed to appreciate the importance
of telling clearly what's wrong with a $\Xi$-like formulation.

In my experience, most philosophers expect the scientists to have some valid --- domain specific --- reasons to
dismiss a $\Xi$-like formulation, but the scientists seem to miss the philosophical motivation for investigating
the precise reasons to exclude a formulation that nobody uses anyway.  As a result, a critical loophole remains
open that prevents a precise identification of the flaws of the most threatening pseudo-scientific theories (and
even prevents to identify a fair amount of ill-conceived academic science).

In fact, the most threatening pseudo-scientific theories follow a pattern that is very close to the one of
Furbetto: (1) they accept most of the state of the art of the best of modern science; (2) but they add also a few
new hypothesis able to: (2a) produce an impressive confirmed prediction (with some luck, but since they are many,
some have luck), (2b) justify the lack of observations of any other unusual phenomena; (2c) predict a new
phenomenon in the future, which is too important to wait for further tests of the theory before acting; (3) they
avoid any discussion about the quality of their new assumptions on the ground that only empirical evidence and
logical coherence matter to evaluate a scientific theory, while the simplicity or ad-hocness of the assumptions are
supposed to be only a matter of taste.  They rarely claim that their theories are {\em better} than the standard
theories: their goal is to get people admit that their theories {\em deserve as much attention as} the standard
ones.

Unfortunately, both scientists and philosophers rarely --- if at all --- challenge pseudo-science on point (3), and
that's why charlatan-scientists do not even need to mention a $\Xi$-like formulation.  But point (3) is actually
the key, because fulfilling points (1) and (2) is always possible, along the lines of Furbetto's example.

We may not need philosophy of science to enable good science, but we definitely nee it to prevent bad or pseudo
science.  And the previous discussion shows that to prevent bad or pseudo science we must give a meaning to
Einstein's {\em ``smallest number of hypothesis''}, in a way that is {\em sufficiently precise} to rule out at
least Furbetto's theory.  By doing this, we will actually get much more.

\section{A well defined framework for science --- what's wrong with Lucky Furbetto's theory}

In this section\footnote{This section is a brief review of the framework introduced in \citep{Scorzato,
    Scorzato-paper2, Scorzato-silfs14}, applied to the case of Furbetto's theory.}, after identifying what's wrong
with the $\Xi$ formulation of Furbetto's theory, I introduce a definition of scientific theories and their
conciseness that essentially justifies the intuitive notion of simplicity commonly used by scientists --- although
often unconsciously.  This provides a clear argument to exclude Furbetto's theory from the circle of scientific
theories, which is also applicable in general. In fact, in Sec.\ref{sec:others}, I apply the same argument to other
less sophisticated but socially more threatening cases of pseudo-scientific theories.

As argued in the Introduction, we must first understand what's wrong with the $\Xi$ formulation of Furbetto's
theory.  Intuitively, it seems implausible that $\Xi$ can be interpreted or {\em measured directly}.  But defining
in general what can or cannot be measured directly is very challenging.  However, as first observed in
\citep{Scorzato}, we can at least identify {\em one necessary property} of any directly measurable quantity, which
is sufficient for our goals:

\begin{post}
\emph{(Directly measurable properties).}
\label{def:ECDM}
The result of a valid {\em direct measurement} of a property $P$ is expressed as: $P=P_0\pm\Delta$, where $P_0$ is
the (unique) central value of the measurement and $\Delta$ the sensitivity of the experimental device (i.e. the
minimal detectable difference between two nearby outcomes).  This expression implies that the probability
distribution for the result of the measurement be centered in $P_0$ and systematically decreases with the distance
from $P_0$ (other features of the distribution are theory dependent).
\end{post}

Postulate~\ref{def:ECDM} might seem obvious and easily fulfilled, but it is not always so.  In particular is it
fulfilled by the variable $\Xi$?  Furbetto can certainly describe entirely the status of the system with just one
variable $\Xi$.  But which precision can he attain in a measurement of $\Xi=\Xi_0\pm\Delta$?  How much is $\Delta$?
Furbetto cannot answer this question. To see why not, note that the interval $\Xi_0\pm\Delta$ should be large
enough to include all the configurations of the system under consideration that differ by an amount that is below
the limits of present experimental sensitivity.  On the other hand, the same interval $\Xi_0\pm\Delta$ should be
small enough to reflect the actual experimental precision, without loss of any information.  This represents a huge
constraint on the possible choice of the formulation $\Xi$.  For example, if we represent a given status of the
system with a G{\"o}del number $\Xi_0$, a slightly different status of the systems will almost always correspond to
a completely different number $\Xi_1$, separated by many numbers that refer to physically very different states
(see \citealp{Scorzato}, for a more general discussion).  In conclusion, although we can represent every system
with a single variable, we have no guarantee that we can (and, indeed, we mostly can't) use that variable to
represent the interval of confidence in which the system might be.

\begin{figure}
\centering
\includegraphics[width=100mm]{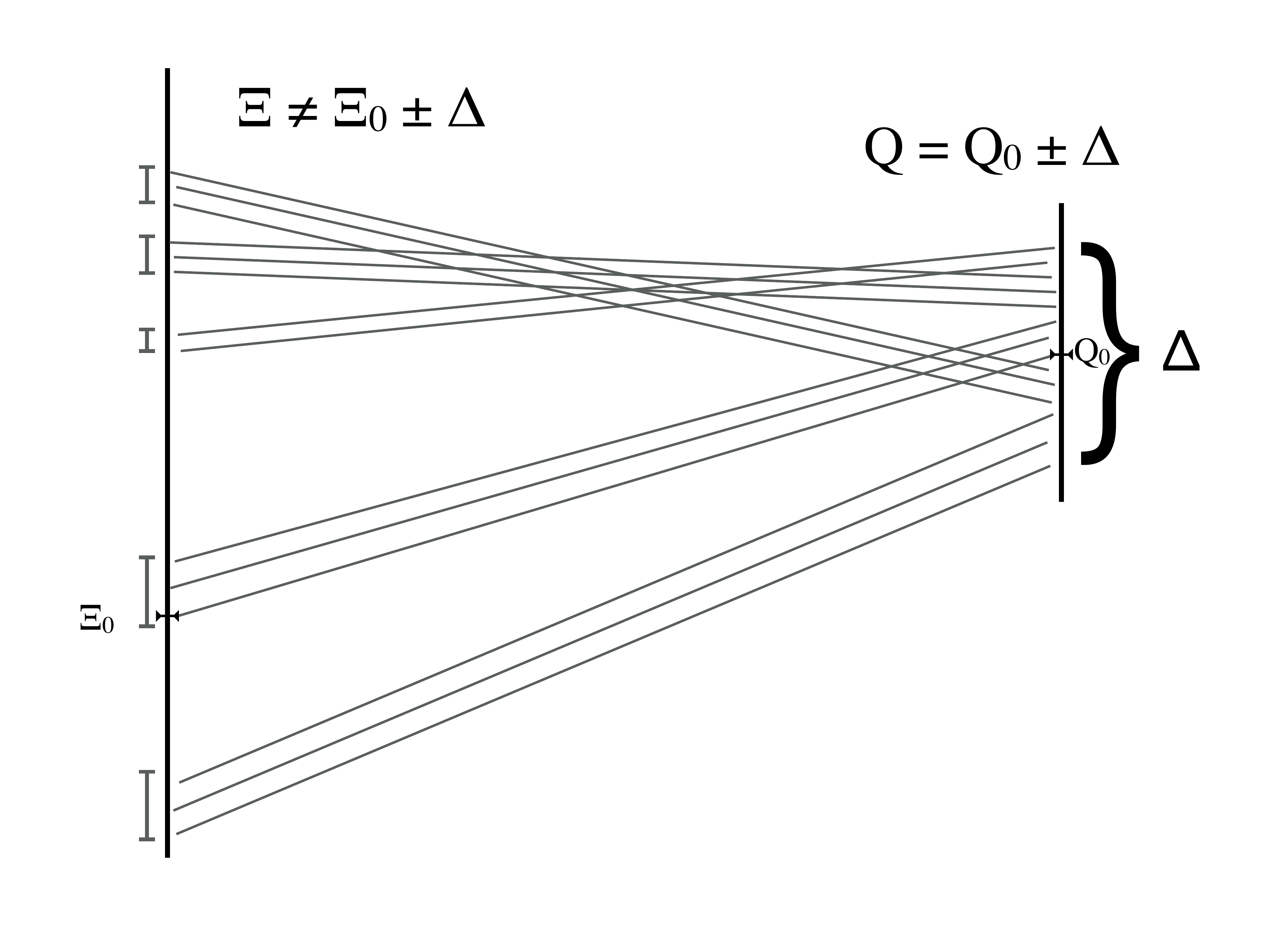}
\caption{The translation between the formulation in terms of $Q$ and the one in terms of $\Xi$ associates a value
  $\Xi_0$ to a value $Q_0$.  But, a connected interval in $Q$ is not necessarily mapped into a connected interval
  in $\Xi$.}
\label{fig:disc}
\end{figure}

Fig.~\ref{fig:disc} illustrates the problem: the variable ($\Xi$) that attains the highest conciseness is not
continuously related to the variable ($Q$) that can be actually measured\footnote{It is tempting to require,
  instead of Postulate~\ref{def:ECDM}, a continuity between $\Xi$ and $Q$.  But this merely shifts the problem of
  characterizing a measurable $\Xi$ into that of characterizing a measurable $Q$.  The way out, adopted in
  Postulate~\ref{def:ECDM}, consists in observing that those properties that lack a continuum relation with any
  property that is practically measurable, they cannot be quoted in terms of a central value and an errorbar.  This
  avoids reference to other properties, besides those that the theory choses as basic measurable.}.  Hence, the
outcome of a measurement cannot be quoted in the form $\Xi=\Xi_0\pm\Delta$.

We have identified a testable limitation of the $\Xi$ formulation of Furbetto's theory.  This is very important,
because it can be used to correct the only serious difficulty of a classic view of scientific theories
\citep{Feigl}.  In fact, we can now describe a scientific theory as a logical-mathematical system whose empirical
interpretation does not rely on mysterious {\em correspondence rules}, but it simply requires that some concepts
are {\em directly measurable}, at least in the testable sense expressed by Postulate~\ref{def:ECDM}.  More
precisely:

\begin{defn}
\emph{(Scientific theories).}
\label{def:ST}
A scientific theory is a quadruple $T=\{P,R,B,L\}$, where 
\begin{itemize}[noitemsep,nolistsep]
\item $P$ is a set of principles\footnote{The principles contain {\em all} the assumptions needed to derive the
  results of the theory, from the logical rules of deduction to the modeling of the experimental devices and of the
  process of human perception, so that no further {\em background science} is needed.  Note that also the {\em
    domain of applicability} of the theory can and must be defined by specifying suitable restrictions in the
  principles themselves.  Any needed auxiliary assumption should also be included.},
\item $R$ is a set of results deduced from $P$ (according to the logic rules included in $P$), 
\item $B$ is a set of properties that appear in $P$ and are {\em directly measurable} in the sense of
  Postulate~\ref{def:ECDM} (we call them Basic Measurable Properties, or BMPs, of $T$),
\item $L$ is the language in which all the previous elements are formulated.
\end{itemize}
\end{defn}

Note that the BMPs do not constitute a {\em universal} basis: there is no such thing.  Note also that
Postulate~\ref{def:ECDM} and Def.~\ref{def:ST} cannot {\em fix} the interpretations of the BMPs.  Nothing can do
that: neither correspondence rules, nor bridge laws, nor models.  Theories face the tribunal of experience as a
whole \citep{Quine2Dogs}, and the {\em assumptions} that their BMPs are sufficiently unambiguous are necessarily
part of the overall theoretical assumptions.  The aim of Postulate~\ref{def:ECDM} is not to {\em fix} the
interpretation of any expression, its aim is rather to {\em exclude} a class of interpretations that we deem
certainly implausible.

Not all the measurable properties of a theory $T$ need to be included in the principles $P$ of $T$. The other
(possibly unlimited) measurable properties can be characterized as follows:

\begin{defn}
\emph{(Measurable properties).}
\label{def:MC}
The measurable properties (MPs) of a theory $T$ are all those properties that can be determined through
observations of the BMPs $B$ of $T$, by possibly employing some results $R$ of $T$.  Their precision is also
determined by $T$.
\end{defn}

Hence, the BMPs must be sufficient to enable the measurements of all the MPs that the theory needs to describe. In
other words, the BMPs provide --- together with the principles to which they belong --- the basis on which the
whole interpretation of the theory is grounded.  Thanks to the identification of the BMPs, the principles truly
encode {\em all the assumptions} of the theory, in a sense that goes beyond the logical structure of the theory,
and it includes the assumptions that {\em we know how to measure some quantities}.  These assumptions might be
wrong, like any other.

A nice model is offered by Euclidean geometry.  The concepts of {\em point} and {\em straight-line} are formally
defined only implicitly through Euclide's postulates.  However, Euclide's mathematical theory becomes a physical
theory once we assume that we know how to recognize a point and a straight-line on a real sheet of paper.  The
precision with which we can recognize a point and a straight-line depends on the thickenss of our pencil and the
imperfections of our ruler.  Any measurement of any other geometrical object ultimately relies on our ability to
recognize precisely points and straight-lines.

Coming to our main example, both Einstein's theory and Furbetto's theory need to assume standard logical rules,
standard mathematics and physics principles.  They also need to assume a model for each experimental device that
they use (we can assume they use the same devices).  They need to assume that the values displayed by the readers
of those devices are unambiguous (these are the BMPs).  The two theories differ because Furbetto's theory assumes a
time dependent $G(t)$, instead of the usual constant $G$; finally, Furbetto's principles include some assumption
that implies (by using also some properties of $G(t)$ that we do not need to detail) the lottery outcome: $(7, 13,
42)$, for July 18th.  The theory that we will compare to Furbetto's theory includes, besides the standard
assumptions of Einstein's theory, also the straightforward assumption that the lottery outcome for July 18th is
$(7, 13, 42)$, with no attempt to establish deeper connections.  In the following, I use the symbol $E$ when
referring to the theory that includes Einstein's assumptions and the straightforward assumption of the correct
lottery result for July 18th, while I use the symbol $F$ for Furbetto's theory.  Hence $F$ and $E$ are fully
equivalent, from the empirical point of view, until next December 31st 2017.

Until now, we haven't yet shown what's wrong with the theory $F$.  Both $E$, $F$ and ``$\Xi=0$'' are valid
theories, according to the previous definitions.  To tell what's wrong with $F$, we first need to tell why $F$ and
``$\Xi=0$'' cannot be two formulations of equivalent theories.  For this we need to define the concept of {\em
  equivalent formulations}, which has a logical aspect and an empirical aspect.  We can make this idea precise with
the following:

\begin{defn}
\emph{(Equivalent formulations for $T$).}
\label{def:EL}
We say that $T$ and $T'$ are equivalent formulations iff:
\begin{itemize}[noitemsep,nolistsep]
\item[(i)] there is a translation ${\cal I}$ between $T$ and $T'$ that preserves the logical structure and the
  theorems ({\em logical equivalence});
\item[(ii)] and for each MP $c$ of $T$ (resp.~$c'$ of $T'$), ${\cal I}(c)$ (resp.~${\cal I}^{-1}(c')$) is also
  measurable with the same precision and the same interpretation (i.e., an experiment that measures $c$ within $T$
  also measures $c'$ within $T'$) ({\em empirical equivalence}).
\end{itemize}
${\cal L}_T$ denotes the set of all pairs $(L,B)$ of available languages and BMPs in which we can reformulate $T$
and obtain a new theory $T'$ that is equivalent to $T$.  In the following, the symbol $T$ refers to a scientific
theory up to equivalent formulations, while $T^{(L[,B])}$ refers to its formulation in the language $L$ [and basis
  $B$].
\end{defn}

Def.~\ref{def:EL} implies, in particular, that the $\Xi$ formulation suggested by Furbetto is not equivalent to $F$.
In fact, the translation that makes them logically equivalent (which does exist!) cannot realize also an empirical
equivalence, because if $\Xi$ stands for all of Furbetto's assumptions, $\Xi$ is not an acceptable BMP and ---
being the only formal property of the theory --- the theory is left with no MP at all.  So, we have been finally
able to draw precisely a useful distinction between valid and non-valid formulations of $F$.

Now that we have made clear why $F$ cannot be expressed as $\Xi=0$, can we also tell precisely what makes $F$
inferior to $E$?  We have already noticed that $E$ offers a simpler set of assumptions to describe all the
phenomena described by $F$.  If we could rely on their respective formulations in ordinary language, we could
clarify what we mean by {\em simpler} merely in terms {\em word counting}.  In fact, $E$ is more concise than $F$,
as already noticed.  Now that we have justified new rules for legitimate re-formulations, that, in particular, rule
out the $\Xi$ formulation, there is no obvious reason why word counting should not be an adequate measure.  So, can
Furbetto still reformulate his theory as concisely as $E$?  No.

In fact, Furbetto's task has become suddenly awfully hard, because adding new symbols won't help him.  To match or
surpass the conciseness of $E$ he should compensate the complexity of his $G(t)$ by relating it more directly to
some measurable quantity.  This is impossible without recognizing a deep empirical meaning for $G(t)$, which is
exactly what his pseudo-scientific theory cannot offer.

Hence, a formulation independent measure of conciseness that emphasizes and exploits these limitations of $F$ can
be defined as the minimal word counting over all legitimate formulations.  More precisely:

\begin{defn}
\emph{(Complexity of the assumptions; conciseness).}
\label{def:CT}
Let $P^{(L,B)}$ be the principles of $T$, when expressed in language $L$ and with BMPs $B$.  Let the {\em
  complexity of the assumptions} of $T$ be:
\begin{equation}
\label{eq:CT}
{\cal C}(T) = \min_{(L,B) \in {\cal L}_T} {\rm length}[P^{(L,B)}]
\end{equation}
Let the {\em conciseness} of $T$ be the inverse of ${\cal C}(T)$: ${\rm Conc}(T)=1/{\cal C}(T)$.
\end{defn}

Note that the measure in Eq.~(\ref{eq:CT}) --- being the minimum over all available equivalent formulations, in the
sense defined previously --- is as language independent as we could possibly wish.  Note, however, that for
theories that include many assumptions from a variety of background theories --- as it is common for any realistic
theories --- it is practically impossible to achieve greater conciseness by employing a radically new language,
without making the connection to measurable quantities even more cumbersome.  This is so, for the same reasons
noted in the case of $F$.  {\em This consideration justifies, for realistic theories, an estimate of conciseness
  based on their formulation in ordinary language}.

In conclusion, theory $F$ should be dismissed because --- even in its most concise formulation available --- it is
unambiguously less concise than another theory ($E$) which is empirically as accurate as $F$.  To be clear, I state
the general criterion as follows:

\begin{defn}
\emph{(theories with no scientific value).}
\label{def:BT}
If two theories $T$ and $T'$ are empirically equivalent, but theory $T'$ is more concise than $T$, then $T$ has no
scientific value\footnote{Note that the determination of conciseness is affected by many sources of uncertainty and
  can only be estimated approximatively \citep{Scorzato-paper2, Scorzato-silfs14}.  Hence, we can say that a theory
  $T$ has no scientific value only within the confidence that we can associate to the determination of its lower
  conciseness with respect to $T'$.  For this reason, any conclusion about the lack of scientific value of a theory
  is necessarily approximate (just like any scientific conclusion), but it is so in a well defined sense of
  approximation.}.
\end{defn}

Note that a theory $T$ may have no scientific value, according to Def.~\ref{def:BT}, for very different reasons: it
may be the result of a completely ill-fated strategy that will never lead to valuable theories, or it might be
simply a failed attempt within a good strategy.  The goal of this criterion is not to give a simple rule to decide
which directions of research are worth pursuing (although it might help also to that purpose), but to give a simple
rule to decide which theories are not worth consideration when we need to relay on their predictions.

Note, finally, that Def.~\ref{def:BT} does not rely on any trade-off between conciseness and empirical accuracy:
conciseness rules out a theory only when it has no empirical advantage.

\section{Other (pseudo-)scientific theories}
\label{sec:others}

In the previous section I have introduced a general and well defined criterion to tell when a theory has no
scientific value, and I have shown that it leads to the expected conclusions for Furbetto's theory.  Does my
criterion work for other (pseudo-)scientific theories?  Since the previous discussion essentially justifies the use
of ordinary language to estimate the complexity of the assumptions of realistic theories, it is not difficult to
apply the same criterion to many other cases.  In this section I consider some notable cases of pseudo-science and
unsuccessful science.  These accounts are necessarily brief, but I think they convey all the important ideas.

It is also natural to ask whether this criterion might also rule out any valuable scientific theory.  I am not
aware of any such case (some examples of valuable theories that are recognized as such are discussed in
\citealp{Scorzato-paper2}).  Finding a counter-example is my challenge to the readers.

It is important to note that the above criterion only compare single theories.  It does not judge entire research
programs.  However, research programs produce many theories, whose value can be compared with those theories that
represent the state of the art.

\subsection{Solipsism}

A paradigmatic case that we should regard as non scientific is {\em solipsism}.  This is actually a very valuable
philosophical idea, because it provides a very good test for any theory of science.  In fact, any good theory of
science should be able to tell why solipsism is not a valuable scientific option.  This is a challenge because
solipsism cannot be excluded neither on logical nor on empirical grounds.  It is sometimes excluded because it is
declared {\em weird}, which is hardly a good philosophical argument.

But the framework of the previous section does give a clear verdict: solipsism requires an unnecessary amount of
assumptions to explain the experience.  In fact, the experiences {\em reported} to the subject by other people
require different explanations --- and hence additional assumptions --- besides those explaining the {\em direct}
experiences of the subject.  What the subject sees and what she hears from the reports of the other people can be
explained much more {\em concisely} by assuming an underlying reality, independent of her mind.  Hence, our rule of
conciseness does the job of declaring solipsism a non scientific option.

\subsection{Intelligent Design}

The theory of Intelligent Design (ID) \citep{dembski2008} has been expressed in many forms, and criticized with
many arguments (see, e.g., \citealp{SoberID,Lutz-ID2}, and references therein, for an introduction to a larger
literature).  But, unfortunately, these criticisms do not address the most careful versions of ID, and they are
therefore too weak.

The careful version of ID that I want to consider here (I call it ID$^*$) is constructed as follows.  ID$^*$
accepts all modern best scientific theories, except that, occasionally --- and precisely in front of complex
biological structures for which there is currently no convincing evolutionary explanation (i.e. those biological
structures that seem to require a very unlikely random mutation to be produced from pre-existing elements) ---
ID$^*$ postulates the intervention of a {\em designer}\footnote{This is also nicely described as the {\em God of
    the gaps} \citep{Coulson}.}.  The existence of such biological structures is admittedly a problem for the
standard theory because, for each puzzling structure, we need to add the (ad-hoc) assumption that a suitable
mutation has indeed occurred, in spite of its low estimated probability.  This is necessary to recover a good
agreement between theory and evidence.

ID$^*$ defies all the philosophical criticisms directed to other forms of ID, because ID$^*$ is impeccable from the
point of view of empirical accuracy, predictivity, falsifiability, testability or similar criteria, for the same
reasons why Furbetto's theory was.  

The sound reason why ID$^*$ is not scientific is another.  Can we express the properties of the {\em designer} in a
way that can be used to deduce the necessary appearance (or at least a higher probability of appearance) of all
those mutations that seem puzzingly unlikely, from the point of view of the standard theory?  If we could do that,
in a way that is more concise than just assuming the appearance of each of them, ID$^*$ would have a point.  But
this is not the case.  The role of the {\em designer}, in any form of ID, is void of any deductive value. If we
clear ID$^*$ from all the rhetoric and leave only the bare minimum which has some deductive value, we find that the
most concise assumption consists in just postulating the appearance of those structure, as the standard theory
does.  This reduces ID$^*$ to something that would not appeal its proponents anymore and it is clearly no better
than the standard theory.

In other words, the fundamental problem of ID is very simple: ID does not even attempt to do what science always
must do: deducing as many phenomena as possible from as few assumptions as possible.  On the contrary, ID de facto
claims that the goal of science is, sometimes, not attainable.  Denying the possibility of scientific progress is
not a crime (and might even be true!), but it has nothing to do with science either.

\subsection{Climate change denials}

Denialism is sometimes defined as the {\em rejection of basic concepts that are undisputed and well-supported parts
  of the scientific consensus}\footnote{See en.wikipedia.org/wiki/Denialism and references therein.}.  Indeed, many
philosophers who analyze single cases of denialism concentrate on assessing the present level of consensus among
the scientific community.  This strategy is inconclusive, at best.  Instead, it is essential to discern good from
bad reasons of dissent, regardless of their popularity.

For example, until about 15 years ago, climate models typically accepted that the effects of anthropogenic $CO_2$
were negligible.  This was justified because the amount of anthropogenic $CO_2$ was much smaller than the amount
routinely exchanged by the oceans.  However, that conclusion could be derived only under further assumptions.
Hence, the fact that anthropogenic $CO_2$ could be neglected was, essentially, a reasonable {\em assumption}.

Around the year 2000 the International Panel for Climate Change (IPCC) started to recognize that those climate
models that {\em calculated} (through computer simulations) the anthropogenic interference on climate were in
better agreement with the data than the models {\em assuming} no interference.  This was a turning point, because
it showed that a more concise model (in which the assumption of negligible anthropogenic effects was
removed\footnote{I assume, as it seems legitimate, that the assumptions needed to use suitable Monte Carlo
  simulations were already needed anyway also in the older models.}) was also in better agreement with the data.
At that point --- with a confidence determined by the associated imprecision --- it became less and less scientific
to assign any value to the old models.

Of course this does not mean that {\em any denial} of significant anthropogenic effect is not scientific.  But if
one does deny it, then he has to indicate {\em which alternative model} he is considering.  Because what is
certainly not scientific (but, sadly, fairly common) is to deny anthropogenic effects while still referring to the
old models, or referring to weaknesses that are present in any model.  A serious scientific statement about some
phenomena should always refer to the model (i.e. the set of assumptions) that it is adopting to deduce those
phenomena.  It is not the denial itself of a phenomenon that can be not scientific, but the choice of the theory to
support the claim.

Unfortunately, the fairly common, but superficial, claims that some scientific results be {\em incontrovertible
  scientific facts} greatly undermine the possibility to draw a robust distinction between scientific and
non-scientific statements.

\subsection{Some versions of multiverses}

Poor science is not reserved to unscrupulous groups determined to fool their audience.  It is a serious risk also
among the best families of scientists.

A typical illusion is that of trading some postulates that we do not like with others that sound more appealing,
but have no scientific advantage.  As long as this exercise is recognized as a simple rephrasing, there is nothing
wrong.  The problem comes when we fall into the illusion of having achieved a much deeper understanding, we
establish research programs and invest resources to elaborate on it.

For example, it is a great temptation to try to {\em explain} the puzzling fundamental constants of the Standard
Model of particle physics and cosmology by saying that, perhaps, there are actually many universes with different
fundamental constants and the one in which we live is simply one of the few that are compatible with
life\footnote{Actually, as far as I know, the only case in which we can convincingly conclude incompatibility with
  life is for a larger cosmological constants.  Most other claims fail to assess all other possible effects.}.

This kind of Multiverse is certainly a possibility.  What is wrong, however, is to present it as an {\em
  explanation}.  Note that we can't naively blame Multiverse for the lack of novel predictions that distinguish it
from the standard theory, more than we can blame the standard theory itself.  The important point is that these
Multiverses also lack of any {\em cognitive} advantage.  But we need a cognitive value like conciseness to
understand it.  In fact, we should ask ourselves: is it simpler to assume some specific constants as fundamental or
is it simpler to assume a more fundamental probability distribution of possible fundamental constants?  There seems
to be little hope that the latter could be more concise, in any realizable formulation.  Hence, we should conclude
that, unless we are able to deduce something more, the advantages of Multiverse are mere illusions.

Once again, it is important to consider only those formulations that don't loose contact with measurable
quantities, and to disregard all the narrative that has no deductive power.

\subsection{The general risk of bad science}

Good science is hard: it requires formulating and testing theories on relevant topics that are more concise and/or
more predictive than the available ones.  This is the only way we have to understand and control better the natural
phenomena.  Often there is no clear cut between bad and good scientific disciplines, but rather a continuum
connecting unsuccessful theories to unsuccessful research programs, hopeless research programs, ill-conceived
research programs, and finally fully fledged pseudo-science.

The temptation to address questions that are meaningful only within a narrow community, because of the loss of
contact with more general goals, is always strong.  This temptation can take many forms: regarding new formulations
as progressive, although they only change the narrative; working for too long on theories that can't even reproduce
our current best theories, working on theories where it is always acceptable to add a new term when we see a
disagreement with the experiments; performing experiments that, if they fail, we can only blame the experiments,
and if they succeed, we won't be able to find a useful application...  It would not be fair to name topics, because
the relevant boundaries are often not between topics, but have a much more complex structure.

An excellent way to avoid all these problems is to be regularly evaluated also by referees who are not field
experts, and hence are less likely to share the same prejudices.  But to do that seriously, it is necessary to
enable them to assess what we have achieved and what we want to achieve.  To this end, it is necessary to recognize
all the assumptions that we are adopting and express them with the simplest possible way, with no decorations.  If
we can't offer a concise, understandable summary to educated non experts, it is unlikely that we are working at the
forefront of science anyway.

\section{Conclusions}

It is often very difficult to tell whether a new theory is valuable: we need to recognize all its explicit and
implicit assumptions, draw its obvious and less obvious consequences, estimate what we might be able to deduce with
more calculations.  We also need to judge whether it represents a fruitful direction of research, beyond its
present, perhaps limited, results.  However, once we have recognized all its assumptions and put together all its
present results, it is usually not difficult at all to determine whether that theory is better than some currently
accepted theory or not.  And it is also not difficult to explain to non-experts the precise reasons of our
conclusion.  Philosophy of science does not need to help scientists to find better theories, but it does need to
identify clearly the general rules behind those theory selections that the scientists regard as obvious.

Although the scientists often feel that they do not need philosophical help for this task, they should recognize
that spelling clearly the detailed rules for their {\em obvious} decision of theory selection is essential to limit
the spread of pseudo-science, to prevent the waste of too many resources on bad science and also the offer better
tools to the society who need to know what the scientists are doing.

In this paper I have argued that it is possible to identify rules that are sufficiently general and precise to tell
what's wrong in many important cases of theories with no scientific value.  But to do this, it was essential to
drop the rhetoric of science as something that {\em proves facts}, and recognize science as the art of finding the
simplest description of what we see.  Moreover, it was essential to recognize that the notion of {\em simplicity}
does have a well defined and rather precise meaning in science, if we use a formulation that refers to measurable
quantities.

This leads to a view of science which is just a small, but crucial, clarification of the famous Einstein's view
already mentioned: {\em The grand aim of all science is to cover the greatest amount of empirical data by logical
  deduction from the smallest amount of hypotheses, quantified as the length of their most concise formulation
  available, that refers to measurable quantities}.  The narrative surrounding some theories may be useful when we
construct them, but not when we assess them: theories must come naked on judgement day.

\bibliography{../philo}{}
\bibliographystyle{chicago}

\end{document}